\documentclass[aps,twocolumn,showpacs,preprintnumbers,superscriptaddress]{revtex4-2}
\usepackage{amsmath}
\usepackage{amssymb}
\usepackage{mathrsfs}
\usepackage{graphicx}
\usepackage{float}
\usepackage{bm}
\usepackage[T1]{fontenc}
\usepackage{natbib}
\usepackage{braket}
\usepackage{epstopdf}
\usepackage{longtable}
\usepackage{multirow}
\usepackage{makecell}
\usepackage{siunitx}
\usepackage[colorlinks=true,citecolor=blue,linkcolor=blue,urlcolor=blue]{hyperref}

\begin{document}

\newcommand{\etal}{{\sl et al.}}
\newcommand{\ie}{{\sl i.e.}}
\newcommand{\sto}{SrTiO$_3$}
\newcommand{\lao}{LaAlO$_3$}
\newcommand{\lno}{LaNiO$_3$}
\newcommand{\otw}{O$^{2-}$}
\newcommand{\alo}{Al$_2$O$_3$}
\newcommand{\aalo}{$\alpha$-Al$_2$O$_3$}
\newcommand{\xto}{$X_2$O$_3$}
\newcommand{\eg}{$e_{g}$}
\newcommand{\tg}{$t_{2g}$}
\newcommand{\dzt}{$d_{z^2}$}
\newcommand{\dxtyt}{$d_{x^2-y^2}$}
\newcommand{\dxy}{$d_{xy}$}
\newcommand{\dxz}{$d_{xz}$}
\newcommand{\dyz}{$d_{yz}$}
\newcommand{\egp}{$e_{g}'$}
\newcommand{\ag}{$a_{1g}$}
\newcommand{\mub}{$\mu_{\rm B}$}
\newcommand{\ef}{$E_{\rm F}$}

\title{High Chern numbers in a perovskite-derived dice lattice (La$X$O$_3$)$_3$/(LaAlO$_3$)$_3$(111) with $X=$ Ti, Mn and Co}


\author{Okan K\"oksal}
\affiliation{Department of Physics and Center for Nanointegration Duisburg-Essen (CENIDE), University of Duisburg-Essen, Lotharstr. 1, 47057 Duisburg, Germany}

\author{L. L. Li}
\affiliation{Department of Physics and Center for Nanointegration Duisburg-Essen (CENIDE), University of Duisburg-Essen, Lotharstr. 1, 47057 Duisburg, Germany}

\author{Rossitza Pentcheva}
\affiliation{Department of Physics and Center for Nanointegration Duisburg-Essen (CENIDE), University of Duisburg-Essen, Lotharstr. 1, 47057 Duisburg, Germany}

\date{\today}

\begin{abstract}

The dice lattice, containing a stack of three triangular lattices, has been proposed to exhibit nontrivial flat bands with nonzero Chern numbers, but unlike the honeycomb lattice it is much less studied. By employing density-functional theory (DFT) calculations with an on-site Coulomb repulsion term, we explore systematically the electronic and topological properties of (La$X$O$_3$)$_3$/(LaAlO$_3$)$_3$(111) superlattices with $X=$ Ti, Mn and Co, where a LaAlO$_3$ trilayer spacer confines the La$X$O$_3$ (L$X$O) dice lattice. In the absence of spin-orbit coupling (SOC) with symmetry constrained to P3, the ferromagnetic (FM) phase of the L$X$O(111) trilayers exhibits a distinct spin-polarized half-metallic state with multiple Dirac crossings and coupled electron-hole pockets around the Fermi energy. Symmetry lowering induces a significant rearrangement of bands and triggers a metal-to-insulator transition. Inclusion of SOC leads to a substantial anomalous Hall conductivity (AHC) around the Fermi energy reaching values up to $\sim-3e^2/h$ for $X=$ Mn and Co in P3 symmetry and both in- and out-of-plane magnetization directions in the first case and along [001] in the latter. The dice lattice emerges as a promising playground to  realise nontrivial topological phases with high Chern numbers.  

\end{abstract}

\maketitle

\section{Introduction}
 
Transition metal oxides (TMO) comprise a class of materials where electronic correlation and the interplay of charge, spin, orbital, and lattice degrees of freedom can lead to fascinating properties ranging from magnetism to superconductivity \cite{TokuraCorrelatedElectronPhysicsTransitionMetal2007,DagottoComplexityStronglyCorrelated2005}. Precise control of the layer thickness, growth orientation, and epitaxial strain of TMO heterostructures provides essential degrees of freedom to tune their functional properties \cite{Roadmap2016}.  Among the TMO heterostructures, perovskite superlattices (SLs) have developed into an excellent platform to explore interface- and confinement-induced phenomena such as interfacial charge transfer, conductivity, magnetism, electronic reconstruction and metal-to-insulator transition \cite{PentchevaElectronicphenomenacomplex2010, HwangEmergentphenomenaoxide2012, ChakhalianColloquiumEmergentproperties2014}. In particular, (111)-oriented SLs have drawn attention due to the possibility to engineer quantum Hall states \cite{XiaoInterfaceengineeringquantum2011}, with view of applications in low-power electronics. Along the [111]- direction  LaO$_3$ and $X$ layers alternate in La$X$O$_3$, thereby, two $X$ triangular lattices form a buckled honeycomb lattice, which is topologically analogous to graphene. Model Hamiltonian studies in conjunction with DFT calculations predicted a distinct set of four bands in (111) bilayers of LaNiO$_3$, two nearly flat interconnected by two dispersive ones with a (Dirac) crossing at $\textrm{K}$ and a quadratic band touching at $\Gamma$ \cite{YangPossibleinteractiondriventopological2011, RueggElectronicstructureLaNiO2012, RueggLatticedistortioneffects2013, DoennigConfinementdriventransitionstopological2014}. Systematic DFT+$U$ calculations have shown that this set of bands occurs also for La$X$O$_3$ ($X=$ Mn and Co \cite{DoennigDesignChernMott2016}) and  a Chern insulator phase driven by spin-orbit coupling (SOC) can emerge in (111) bilayers of La$X$O$_3$ ($X=$ Mn \cite{WengTopologicalmagneticphase2015, DoennigDesignChernMott2016}, Co \cite{DoennigDesignChernMott2016}, Ru, Os \cite{Guo-npjQM}, Pd, and Pt \cite{LuStrainonsitecorrelationtunable2019, KoksalCherntopologicalinsulating2019}). The (111)-oriented LaMnO$_3$ buckled honeycomb bilayer was predicted to have a nontrivial band gap of $\sim$ 150 meV and a quantized Hall conductivity of $-e^2/h$ \cite{DoennigDesignChernMott2016}. On the other hand, the Chern insulating phases are often unstable with respect to symmetry breaking that leads to trivial Mott insuling ground states, albeit with properties distinct from the bulk compounds \cite{DoennigDesignChernMott2016}. In the meantime, the growth of (111)-oriented superlattices has been successfully demonstrated, thus enabling the exploration of such exotic phases \cite{HerranzHighmobilityconduction2012, MiddeyEpitaxialgrowth1112012, PiamontezeInterfacialpropertiesmathrmLaMnO2015, WeiFerromagneticphasetransition2017, ArabNanolett2019, Longlong2022}. Recent advances in the theoretical understanding, fabrication, and characterization of correlated and topological phases in (111)-oriented perovskite-derived heterostructures have been highlighted in Ref.~\cite{ChakhalianStronglycorrelatedtopological2020}. 

\begin{figure}[htbp]
\centering
\includegraphics[width=0.4\textwidth]{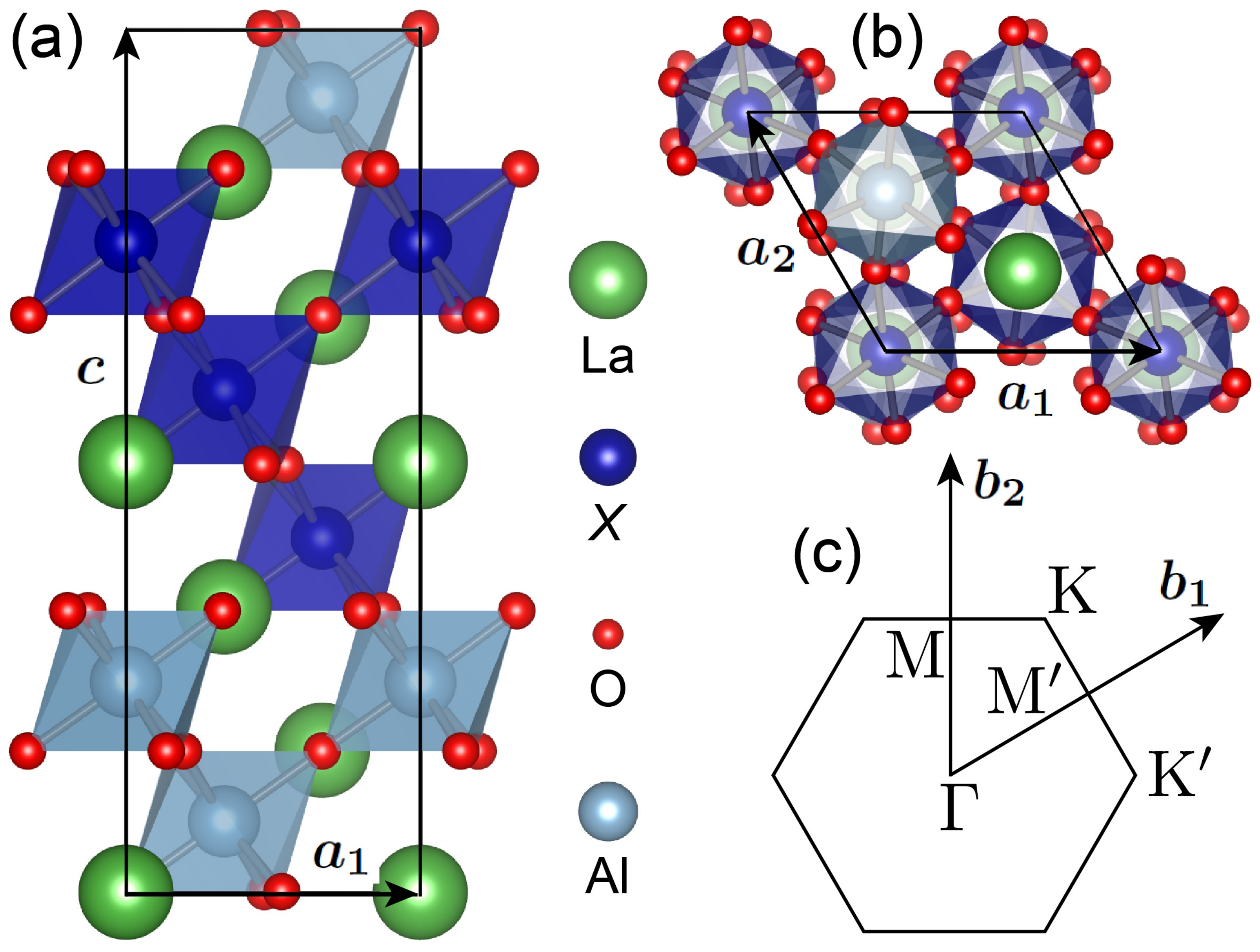}
\caption{(a) and (b) Side and top views of a (La$X$O$_3$)$_3$/(LaAlO$_3$)$_3$(111) superlattice ($X=$ Ti, Mn and Co), where the La$X$O$_3$ trilayer forms a dice lattice that is confined to the LaAlO$_3$ trilayer. In this dice lattice one of the three $X$ sites is located in the central layer and the other two on the interface layers. 
\label{fig1}$\boldsymbol{a}_1$ and $\boldsymbol{a}_2$ are the lateral lattice vectors of the dice lattice and $\boldsymbol{c}$ is the out-of-plane lattice vector. (c) Two-dimensional Brillouin zone of the dice lattice with $\boldsymbol{b}_1$, $\boldsymbol{b}_2$ are the reciprocal lattice vectors of the real-space counterparts $\boldsymbol{a}_1$, $\boldsymbol{a}_2$ while $\Gamma$, $\textrm{K}$, $\textrm{K}'$, $\textrm{M}$, $\textrm{M}'$ denote the high-symmetry k-points.}
\end{figure}

In contrast to the honeycomb bilayers in (111)-oriented perovskites, less attention has been paid to their trilayer counterparts. In the latter, a stack of three triangular lattices forms a so-called \textit{dice} lattice. Tight-binding Hamiltonian studies have shown that the dice lattice and its one-dimensional ribbon structure exhibit nontrivial electronic properties expressed in terms of in-gap flat bands, nonzero Chern numbers, quantum anomalous Hall conductance, and chiral edge states \cite{WangNearlyflatband2011, SoniFlatbandsferrimagnetic2020}. In particular, a nearest-neighbor tight-binding model on a bipartite lattice with distinct number of neighbors for the outer vs. inner layer and  Rashba-type SOC  predicted nearly flat bands  with a Chern number $C\,=\,\pm2$ \cite{WangNearlyflatband2011}. 
Material-specific DFT calculations for the dice lattice are rare: A half-metallic phase was identified in the (111) trilayers of LaNiO$_3$, with the interfacial Ni $e_g$ states contributing to a spin-polarized conduction \cite{DoennigConfinementdriventransitionstopological2014}. To assess the possibly nontrivial properties of the dice lattice, in this work we explore  (La$X$O$_3$)$_3$/(LaAlO$_3$)$_3$(111) SLs with $X=$ Ti, Mn and Co, where a LaAlO$_3$ trilayer spacer confines the La$X$O$_3$ dice lattice, as displayed in Fig. \ref{fig1}. While the model of Wang and Ran assumed an $s$-orbital~\cite{WangNearlyflatband2011}, here we consider a $d^1$ configuration for $X=$ Ti (\tg$^1$) and an \eg$^1$ for $X=$ Mn and Co. Such an orbital configuration has been found essential in order to achieve topologically nontrivial behavior for the honeycomb layers~\cite{DoennigDesignChernMott2016}.  We study systematically the electronic and topological properties of the considered SLs by performing DFT calculations with an on-site Coulomb repulsion term. In particular, we investigate how the interplay of ferromagnetism, SOC, and lattice symmetry influence the band structure, Berry curvature, and anomalous Hall conductivity of (La$X$O$_3$)$_3$/(LaAlO$_3$)$_3$(111) SLs. Parallels to the honeycomb counterpart (La$X$O$_3$)$_2$/(LaAlO$_3$)$_4$(111) SLs \cite{DoennigDesignChernMott2016} are also discussed.

\section{Theoretical Approach}

Systematic DFT calculations were performed for (La$X$O$_3$)$_3$/(LaAlO$_3$)$_3$(111) SLs ($X=$ Ti, Mn and Co) with 30 atoms in the primitive cell using the projector augmented wave method, \cite{Kresseultrasoftpseudopotentialsprojector1999} as implemented in the VASP code \cite{KresseEfficientiterativeschemes1996}. The generalized gradient approximation (GGA) was used for the exchange-correlation functional, as parameterized by Perdew, Burke, and Enzerhof \cite{PerdewGeneralizedGradientApproximation1996}. Static correlation effects were included in the GGA$+U$ approach \cite{DudarevElectronenergylossspectrastructural1998} by employing $U=5$ eV for the $X$ $3d$ orbitals and $U=8$ eV for the La $4f$ orbitals, in line with previous work \cite{WengTopologicalmagneticphase2015, DoennigDesignChernMott2016}. A detailed examination of the topological properties for $X=$ Mn as a function of the $U$ parameter is provided in the Supplemental Material, showing that the topological phases are robust for variation of $U$ between 3.5 and 6.0 eV. A cutoff energy of 600 eV was used to truncate the plane-wave expansion and a $\Gamma$-centered $k$-point mesh of $8\times8\times2$ to sample the Brillouin zone (BZ). In order to model the SLs grown on a LaAlO$_3$(111) substrate, the lateral lattice constant was set to $a_{\textrm{LAO}} \times \sqrt{2}$, ($a_{\textrm{LAO}}=3.79$ {\AA}). The out-of-plane lattice parameter and the internal coordinates were optimized until the forces on atoms were less than 0.01 eV/{\AA} and the change in total energy was less than $10^{-6}$ eV. Octahedral rotations and distortions were fully taken into account in the structural optimization. The Fermi surfaces were calculated using wannier90 \cite{Mostofiwannier90toolobtaining2008, Mostofiupdatedversionwannier902014} and plotted using Fermisurfer \cite{KawamuraFermiSurferFermisurfaceviewer2019}. Spin-orbit coupling (SOC) was included with magnetization direction parallel or perpendicular to the [111]-direction. For the topological analysis, maximally localized Wannier functions (MLWFs) \cite{MarzariMaximallylocalizedWannier2012} were constructed to compute the Berry curvature and anomalous Hall conductivity of (La$X$O$_3$)$_3$/(LaAlO$_3$)$_3$(111) SLs on a dense $k$-point mesh of $144 \times 144 \times 12$ using the wannier90 code \cite{Mostofiwannier90toolobtaining2008, Mostofiupdatedversionwannier902014}.

\section{Results and Discussion}

In the following, we consider the electronic and magnetic properties of (La$X$O$_3$)$_3$/(LaAlO$_3$)$_3$(111) SLs ($X=$ Ti, Mn and Co) with ferromagnetic (FM) order, which was found to be more stable than a layerwise antiferromagnetic (AFM) arrangement (see Supplemental Material).

\begin{figure*}[htbp]
\centering
\includegraphics[width=1.0\textwidth]{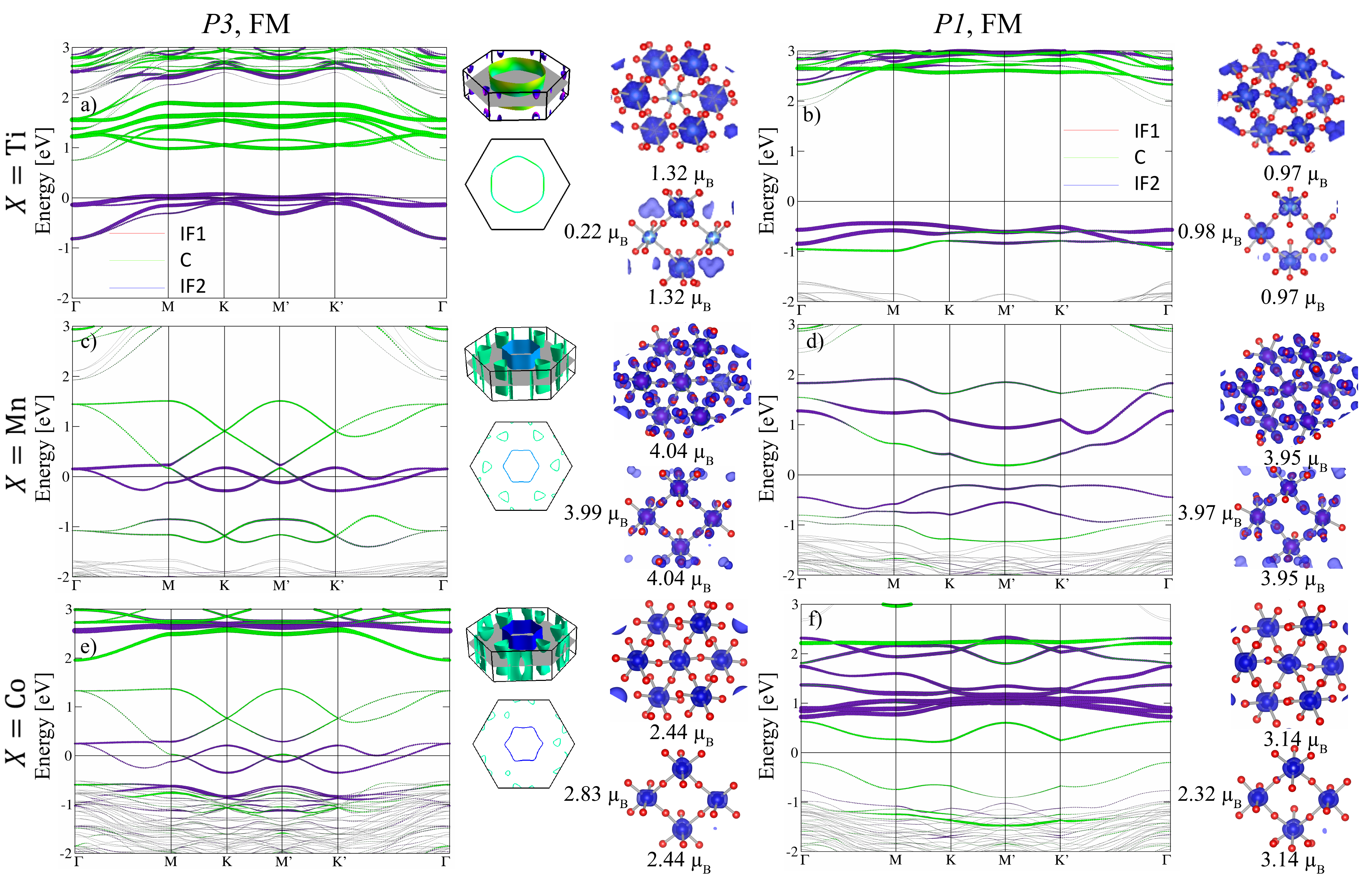}
\caption{Spin-resolved/site-projected band structures, band-decomposed Fermi surfaces, and top/side-view spin densities of (La$X$O$_3$)$_3$/(LaAlO$_3$)$_3$ (111) SLs: (a-b) for $X=$ Ti, (c-d) for $X=$ Mn and (e-f) for $X=$ Co with the isosurface values of 0.01 $e$/{\AA}$^3$ for $X=$ Ti, Mn and 0.05 $e$/{\AA}$^3$ for $X=$ Co. The band structures and the spin densities are shown for both P3 (left) and P1 (right) symmetries. In the band structures, color/black curves represent the majority/minority bands. The Fermi level is set to zero and denoted by a dashed line. Red, green, and blue colors represent the contributions from the first interface (IF1), the central (C), and the second interface (IF2) $X$ layers, respectively. Side and top view of the Fermi surfaces are shown with electron pockets in purple/blue and hole pockets in green. The spin densities were integrated in the energy range between --8 eV to $E_{\rm F}$ except for c-d) where the integration interval is from --1.3 eV to $E_{\rm F}$. We show also the magnetic moments on the $X$ sites  in units of $\mu_B$.}
\label{figure2}
\end{figure*}

\subsection{GGA+$U$ results}

\begin{table*}

\caption{Structural, magnetic, and electronic properties of ferromagnetic (La$X$O$_3$)$_3$/(LaAlO$_3$)$_3$(111) SLs ($X=$ Ti, Mn and Co) for P3 and P1 symmetries in the absence of SOC. $\Delta E$ is the energy difference of the system in P3 symmetry compared to P1, $c$ is the optimized out-of-plane lattice constant, $d_{X\textrm{-O}}$ (IF/C) the $X$-O bond lengths in \AA\ in the interfacial/central $X$O$_6$ octahedra, $E_g$ the band gap in eV, and $M_S$ the layer-resolved spin magnetic moments (in units of $\mu_B$) at the $X$ sites, IF1/IF2 denote the first/second interface layer and C the central one.}\label{tab1}

\begin{ruledtabular}
\begin{tabular}{cccccccc}

$X$ & Symmetry  & $\Delta E$ (eV) & $c$ ({\AA}) & $d_{X\textrm{-O}}$ (IF)   & $d_{X\textrm{-O}}$ (C) & $E_g$ (eV) & $M_S$ (IF1/C/IF2) \\

\hline

\multirow{2}{*}{Ti}
& P3  &2.57 & 14.23  &2.06-2.07 & 1.96 & Metal & 1.32/0.22/1.32  \\
& P1  &0    & 14.34  &2.01-2.14 & 2.04-2.09 & 2.34 & 0.97/0.98/0.97 \\	

\hline

\multirow{2}{*}{Mn}
& P3  &0.64 & 13.78  & 1.96-2.00  & 1.96 & Metal & 4.04/3.99/4.04 \\
& P1  &0    & 14.00  & 1.89-2.12 & 1.93-2.10 & 0.40 & 3.95/3.97/3.95 \\

\hline

\multirow{2}{*}{Co}
& P3  &0.55 & 13.58 & 1.97-1.98 & 1.91 & Metal & 2.44/2.83/2.44 \\
& P1  &0    & 13.77 & 1.87-2.08 & 1.92-2.08 & 0.44 & 3.14/2.32/3.14  \\

\end{tabular}
\end{ruledtabular}
\end{table*}

To determine the most stable configuration we have considered both SLs constrained to (P3) symmetry as well as released constraints, leading to (P1) symmetry. Table \ref{tab1} lists the structural, magnetic, and electronic properties of (La$X$O$_3$)$_3$/(LaAlO$_3$)$_3$(111) SLs ($X=$ Ti, Mn and Co) for both P3 and P1 symmetries in the absence of SOC. For $X=$ Ti, Mn and Co, P1 symmetry is energetically favored over P3 by $\sim2.57\,(1.57)$, $0.64\,(0.64)$, and $0.55\,(0.56)$ eV per 30-atom unit cell, respectively, where the values in the brackets indicate the energy differences including spin-orbit coupling. The reduction of symmetry is accompanied by an expansion of the out-of-plane lattice constant $c$ from 14.23 {\AA} (P3) to 14.34 {\AA} (P1) for $X=$ Ti, from  13.78 {\AA} (P3) to 14.00 {\AA} (P1)  for $X=$ Mn and from 13.58 {\AA} (P3) to 13.77 {\AA} (P1) for $X=$ Co. Furthermore, the reduction of symmetry leads to a strong variation of the $X$-O bond lengths. The shortest Ti-O bonds are obtained for the central layer (1.96 \AA) in P3 symmetry, the ones in the interface layer being  ($\sim$2.07 \AA). In contrast, a strong bond variation occurs for P1 symmetry in the interface layers (2.01-2.14 \AA) compared to the central layer (2.04-2.09 \AA). For $X=$ Mn the bond lengths lie in a narrow range  (1.96-2.00 \AA) for P3 symmetry, but the disparity is enhanced to 1.89-2.12 {\AA} in the interface layers and to 1.93-2.10 {\AA} in the central (C) layer for P1 symmetry, indicating a strong Jahn-Teller (JT) effect. Similarly, the Co-O bonds change from 1.97-1.98 \AA\ for P3 symmetry to 1.87-2.08 \AA\ (IF1/IF2) and 1.92-2.08~\AA\ (C) for P1 symmetry. 


The structural changes are closely related to the electronic properties. Fig. \ref{figure2} displays the spin-resolved, site-projected band structures, the band-decomposed Fermi surfaces, and the top/side-view of the spin densities of (La$X$O$_3$)$_3$/(LaAlO$_3$)$_3$(111) SLs for both P3 and P1 symmetries. In all cases the band structure close to the Fermi level is dominated by majority spin bands, leading to halfmetallic behavior for P3 symmetry. We first discuss the high symmetry cases (left panels). For $X$ = Ti the asymmetry of the interface vs. central layer for the P3 case discussed above is reflected also in the band structure (Fig. \ref{figure2}a):  while the $3d$ bands of the central lie above 1~eV (green), the four bands around the Fermi level have exclusively interface character (purple) with two more dispersive just below \ef\ and two relatively flat at the Fermi level, both sets touching at K  and $\textrm{K}'$. 

The band structures for $X=$ Mn and Co (Fig. \ref{figure2}c, e) with P3 symmetry show similar features and are dominated  by majority bands with multiple Dirac crossings around the Fermi level as well as quadratic band touching at $\Gamma$ slightly above the Fermi energy and at $\sim 1.5$ eV. These spin-polarized bands are grouped into three distinct pairs: The lowest occupied and the highest unoccupied pairs of bands are predominantly localized at the central layer (green), whereas the middle pair of bands at the Fermi level is of prevailing interface character (purple). Both the top and bottom pairs of bands (for Co the latter overlaps with the O $2p$ bands) show two Dirac crossings which reside at $\textrm{K}$ and $\textrm{K}'$. The middle pair intersects the Fermi level leading to a half-metallic behavior and exhibits three crossings, located along $\textrm{M}$-$\textrm{K}$, $\textrm{K}$-$\textrm{M}'$ and $\textrm{M}'$-$\textrm{K}'$.  

As seen from the site-projected band structures (Fig. \ref{figure2}a, c, e), the Fermi surface is dominated by bands of the interfacial layers (purple colors), whereas the bands of the central layer (green color) are shifted from the Fermi energy. This indicates that within P3 symmetry electron conductivity in (La$X$O$_3$)$_3$/(LaAlO$_3$)$_3$(111) SLs is mainly contributed by the $3d$ bands of the interface $X$-ions. Due to the halfmetallic nature and intertwined bands around $E_{\rm F}$, the Fermi surface contains coupled electron-hole pockets, which exhibit different features for $X=$ Ti, Mn and Co (Fig. \ref{figure2}): six electron pockets (purple) around K and one hole pocket (green) around $\Gamma$ for $X=$ Mn and Co and v.v. for Ti.

\begin{table*}

\caption{Electronic and magnetic properties of (La$X$O$_3$)$_3$/(LaAlO$_3$)$_3$(111) SLs ($X=$ Ti, Mn and Co) for P3 and P1 symmetries including SOC with magnetization direction along the [100] and [001] directions. $\Delta E = E_{[100]}-E_{[001]}$ is the magnetocrystalline anisotropy energy (MAE), $E_g$ the band gap, $M_S$ and $M_L$ are the spin and orbital moments (in units of $\mu_B$).}

\label{tab2}
\begin{ruledtabular}
\begin{tabular}{ccccccc}

$X$ & Symmetry & SOC & $\Delta E $ (meV) & $E_g$ (eV) & $M_S$ (IF1/C/IF2) & $M_L$ (IF1/C/IF2) \\

\hline

\multirow{4}{*}{Ti}
& \multirow{2}{*}{P$_3$} & [100] &0.0 & 0.83 & 1.03/0.92/1.02 & -0.04/-0.01/-0.04 \\
&  & [001] &0.07 & 0.83 & 1.04/0.93/1.02 & -0.06/-0.03/-0.06 \\
\cline{2-7}
& \multirow{2}{*}{P$_1$} & [100] &0.18 & 2.33 & 0.97/0.98/0.99 & -0.01/-0.03/-0.01 \\
&  & [001] &0.0  & 2.33 & 0.98/0.99/0.99 & -0.06/-0.07/-0.06 \\

\hline

\multirow{4}{*}{Mn}
& \multirow{2}{*}{P$_3$} & [100] &0.0 & Metal & 3.99/3.99/3.99 & -0.02/-0.01/-0.01 \\
&  & [001] &0.34 & Metal & 3.99/3.98/4.0 & -0.01/-0.01/-0.01 \\
\cline{2-7}
& \multirow{2}{*}{P$_1$} & [100] & 0.02 & 0.39 & 3.91/4.0/3.91 & -0.01/-0.00/-0.01 \\
&  & [001] &0.0  & 0.39 & 3.91/4.01/3.91 & -0.00/-0.02/-0.00 \\

\hline

\multirow{4}{*}{Co}
& \multirow{2}{*}{P$_3$} & [100] &0.0 & Metal & 2.44/2.82/2.45 & 0.23/0.21/0.23 \\
&  & [001] &4.79  & Metal & 2.43/2.83/2.43 & 0.16/0.07/0.16 \\
\cline{2-7}
& \multirow{2}{*}{P$_1$} & [100] &2.38 & 0.42 & 3.04/2.36/3.04 & 0.11/0.24/0.11 \\
&  & [001] &0.0  & 0.42 & 3.35/1.96/3.35 & 0.21/0.16/0.21 \\

\end{tabular}
\end{ruledtabular}
\end{table*}

The symmetry lowering and structural distortion from P3 to P1  triggers a significant reconstruction of the band structures of $X=$ Ti, Mn and Co (Fig. \ref{figure2}b, d, f) leading in all cases to a metal-to-insulator transition. The degeneracy of the $\textrm{M}$ and $\textrm{M}'$ points is lifted and the set of spin-polarized bands around the Fermi energy splits into an occupied valence and an empty conduction band separated by a substantial band gap of 2.34, 0.40, and 0.44 eV for $X=$ Ti, Mn and Co (cf. Table \ref{tab1}), respectively. For $X=$ Ti three narrow bands are occupied just below $E_{\rm F}$, the lowest one with predominant central-layer contribution and the other two of prevailing interface character. The highest valence band of $X=$ Mn has interface character, whereas the conduction band is of mixed character with its bottom at $\textrm{M}'$ having a stronger contribution from the central layer. On the other hand for  $X=$ Co both the top valence and bottom conduction band  have a predominant contribution from the central layer and the interface bands lie away from $E_{\rm F}$.

 The spin densities and magnetic moments in the IF1/C/IF2 layers for P3 and P1 symmetries, shown in Fig. \ref{figure2}, the latter listed also in Table \ref{tab1}, provide further insights into the electronic reconstruction.  The pronounced asymmetry between the central (C) and interface (IF1 and IF2) layers for $X=$ Ti in P3 symmetry  is also reflected in the magnetic moments 1.32/0.22/1.32 $\mu_{\rm B}$. This indicates a modulation of the Ti valence state: Ti$^{3\pm\delta}$ in the interface and Ti$^{4+}$ in the central layer, which is  consistent with the band occupation pattern described above (Fig. \ref{figure2}a): a fully and a partially occupied band for each interface layer and empty $d$ bands for the central layer. In the (111)-oriented SLs the octahedral symmetry is reduced to trigonal which splits the $t_{2g}$ triplet into an $a_{1g}$ singlet and an $e'_g$ doublet. The spin density for P3 symmetry (Fig. \ref{figure2}) displays a preferential occupation of the $e'_g$ doublet in the interface layer. In contrast, for P1 symmetry similar magnetic moments in all layers 0.97/0.98/0.97 $\mu_{\rm B}$ are obtained, consistent with a valence state of Ti$^{3+}$ in all layers. Interestingly, the spin density for P1 symmetry (Fig. \ref{figure2}) reflects a staggered orbital polarization of $d_{xz}$ and $d_{yz}$ orbitals instead of the expected $a_{1g}$ or $e'_{g}$ orbitals. This is similar to the behavior found in the honeycomb (LaTiO$_3$)$_2$/(LaAlO$_3$)$_4$(111) in P1 symmetry \cite{DoennigDesignChernMott2016}. 
 
 For $X=$ Mn (Fig. \ref{figure2}c, d) the magnetic moments are almost unchanged between the outer and inner layer and between P3 (4.04/3.99/4.04 $\mu_B$) and P1 symmetry (3.95/3.97/3.95 $\mu_{\rm B}$) and are consistent with a high spin Mn$^{3+}$  $d^4$ configuration. Moreover, the metal-to-insulator transition from P3 to P1 symmetry can be understood as a result of a Jahn-Teller (JT) distortion of the JT active Mn$^{3+}$ ion, analogous to the behavior observed in the honeycomb Mn-bilayer \cite{DoennigDesignChernMott2016}. For $X=$ Co$^{3+}$ in P3 symmetry (Fig. \ref{figure2}), the magnetic moment at the Co sites in the interface layers is $2.44$ $\mu_B$ and the central layer acquires a magnetic moment of $2.83$ $\mu_B$, indicating an intermediate-spin state ($t_{2g}^5$, $e_{1g}^{1}$) (Fig. \ref{figure2}). In contrast, for P1 symmetry the sizes of magnetic moments are reversed with enhanced magnetic moments in the interfacial layer (3.14 $\mu_{\rm B}$) rather pointing to a high-spin state, while the magnetic moment in the central layer is nearly 1~$\mu_{\rm B}$ smaller (2.32 $\mu_{\rm B}$). We note that different spin states of Co were also reported for the honeycomb Co-bilayers \cite{DoennigDesignChernMott2016} and are related to the rich phase diagram of bulk LaCoO$_3$ with respect to spin degree of freedom, e.g., with transitions from a low-spin ($t_{2g}^6$) ground state to a ferromagnetic intermediate- or high-spin state e.g. under pressure or  strain \cite{HsuFerromagneticinsulatingstate2012,Geisler2020}.

\subsection{Effect of SOC and Topological Analysis}

In the following, we proceed with the effect of SOC and topological analysis in (La$X$O$_3$)$_3$/(LaAlO$_3$)$_3$(111) SLs with $X=$ Ti, Mn and Co. For the analysis of the topological properties, we performed a Wannier interpolation of the relevant part of the DFT+$U$+SOC band structure around the Fermi energy with prevailing $X$ $3d$ character and calculated both the Berry curvature and anomalous Hall conductivity (AHC) for the low-energy $X$ $3d$ bands by constructing the MLWFs \cite{MarzariMaximallylocalizedWannier2012} using the wannier90 code \cite{Mostofiwannier90toolobtaining2008, Mostofiupdatedversionwannier902014}. The quality of the Wannier fit is demonstrated in the Supplemental Material by superimposing the Wannier interpolated bands on the original DFT bands for $X=$ Ti, Mn and Co.

\begin{figure*}[htbp]
\centering
\includegraphics[width=0.8\textwidth]{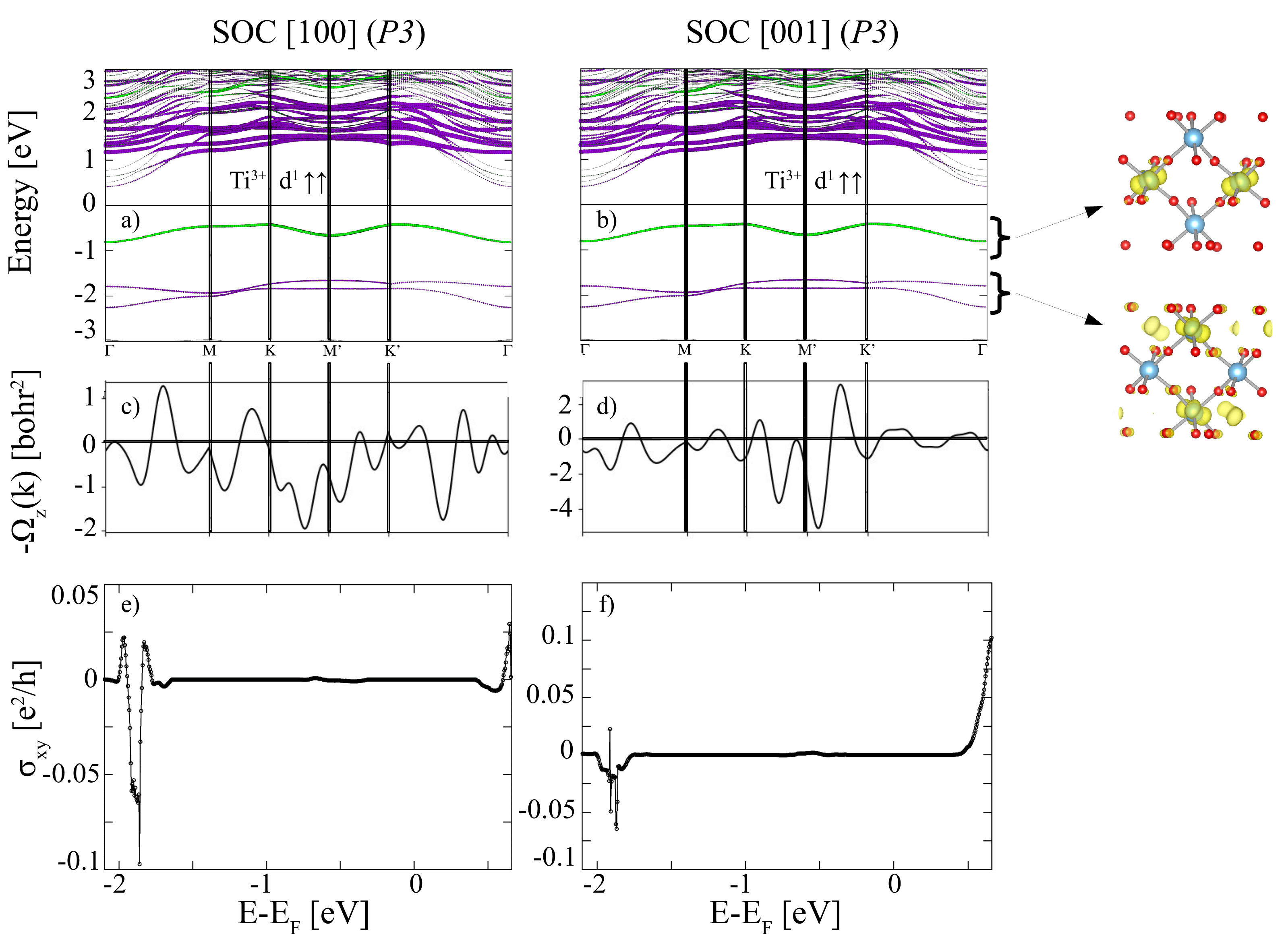}
\caption{(a-b) Element-projected GGA+$U$+SOC band structures with the same color coding as used in Fig. \ref{figure2} for $X=$ Ti with P3 symmetry and magnetization directions along [100] and [001]. Additionally, the electron density distribution  integrated in the energy range between --2.3 and --1.5 eV, as well as --1.0 and --0.3 eV at the isosurface value of 0.01 $e$/{\AA}$^3$ is shown; (c-d) Berry curvatures $\Omega_{xy}(\textbf{k})$ along the same $\textit{k}$-path; (e-f) the corresponding anomalous Hall conductivities $\sigma_{xy}^{\textrm{AHC}}$ vs. the chemical potential in units of $e^{2}/h$. }
\label{figure3}
\end{figure*}

\begin{figure*}[htbp]
\centering
\includegraphics[width=0.8\textwidth]{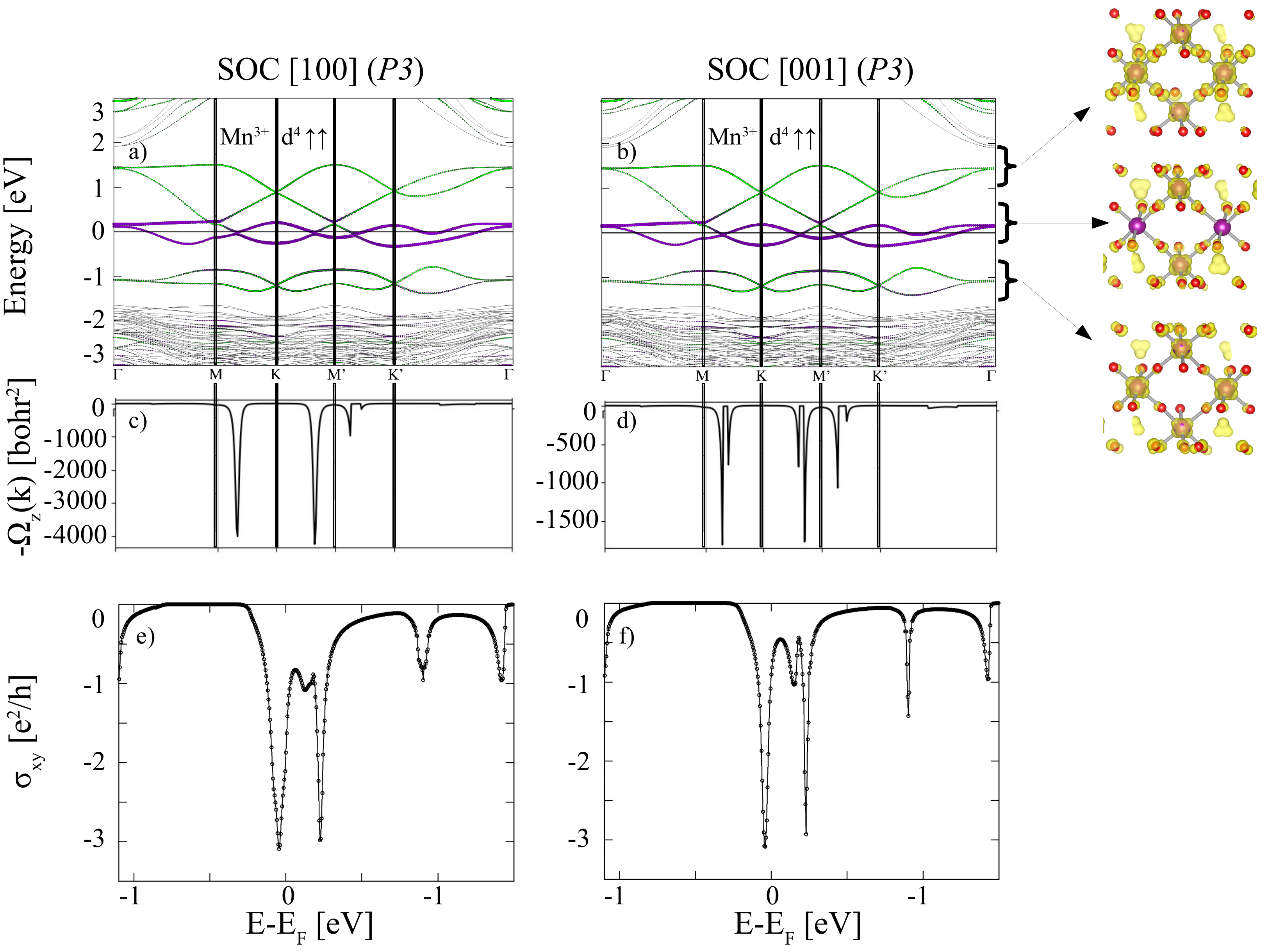}
\caption{(a-b) Element-projected GGA+$U$+SOC band structures with the same color coding as used in Fig. \ref{figure2} for $X=$ Mn with P3 symmetry and magnetization directions along [100] and [001]. Additionally, the electron density distribution of integrated in the energy range between --1.4 and --0.7 eV, --0.3 and 0.22 eV and 0.23 and 1.55 eV at the isosurface value of 0.01 $e$/{\AA}$^3$ are displayed; (c-d) corresponding  Berry curvatures $\Omega_{xy}(\textbf{k})$ along the same $\textit{k}$-path; (e-f) anomalous Hall conductivities $\sigma_{xy}^{\textrm{AHC}}$ vs. the chemical potential in units of $e^{2}/h$. }
\label{figure4}
\end{figure*}

The Berry curvature is calculated using the Kubo formula \cite{Wanginitiocalculationanomalous2006,YatesSpectralFermisurface2007}

\begin{equation}\label{berrykubo}
 \Omega_{xy}^z(\textbf{k})= -2\sum_{n \in occ} \sum_{m\neq n} \textrm{Im} \frac{ \left\langle \psi_{n\textbf{k}} \left| v_{x} \right| \psi_{m\textbf{k}} \right\rangle
\left\langle \psi_{m\textbf{k}} \left| v_{y} \right| \psi_{n\textbf{k}} \right\rangle}{{(\epsilon_{m\textbf{k}}-{\epsilon}_{n\textbf{k}})}^{2}},
\end{equation}

where the sum over $n$ is restricted to the occupied bands, $\psi_{n\textbf{k}}$ is the spinor wave function of the $n^{th}$ band, $\epsilon_{n\textbf{k}}$ is the corresponding band energy, and $v_x$ ($v_y$) is the velocity operator along the $x$ ($y$) direction. The anomalous Hall conductivity is calculated by integrating the Berry curvature over the Brillouin zone (BZ) which is transformed into a sum over $\textbf{k}$-points,

\begin{equation}\label{ahc}
\sigma_{xy}^{\textrm{AHC}}= -\frac{e^{2}}{\hbar} \frac{1}{N_{\textbf{k}}V_c}\sum_{\textbf{k}} \Omega_{xy}^z(\textbf{k}),
\end{equation}

where $V_c$ is the volume of the unit cell and $N_{\textbf{k}}$ the number of $\textbf{k}$-points used for sampling the BZ.

\begin{figure*}[htbp]
\centering
\includegraphics[width=0.8\textwidth]{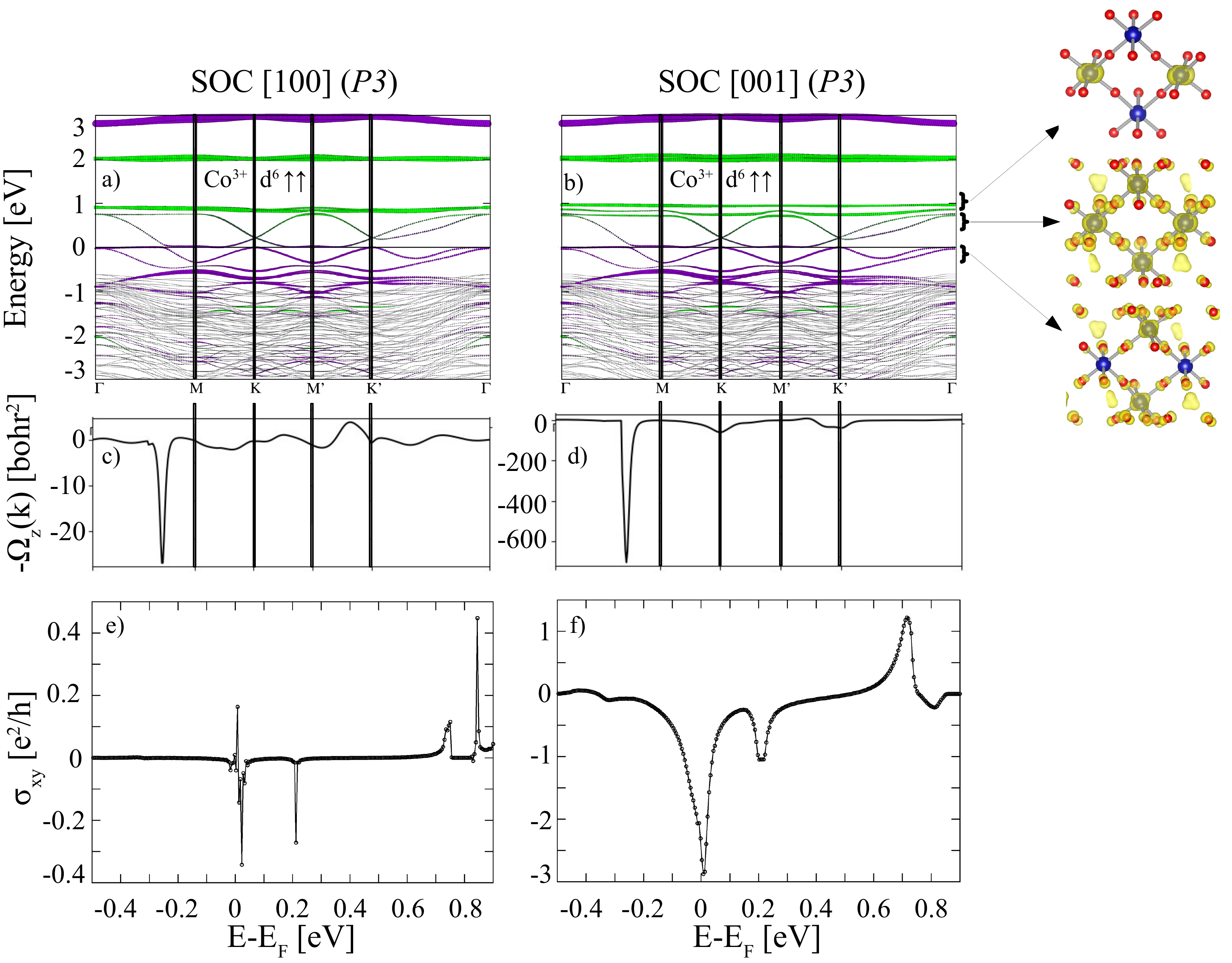}
\caption{(a-b) Element-projected GGA+$U$+SOC band structures with the same color coding as used in Fig. \ref{figure2} for $X=$ Co with P3 symmetry and magnetization directions along [100] and [001]. Additionally the  electron density distribution integrated in the energy range between --0.5 and \ef, \ef\ and 0.74 eV and 0.8 and 1.0 eV at the isosurface value of 0.01 $e$/{\AA}$^3$ is displayed; (c-d) Berry curvatures $\Omega_{xy}(\textbf{k})$ along the same $\textit{k}$-path; (e-f) corresponding anomalous Hall conductivities $\sigma_{xy}^{\textrm{AHC}}$ vs. the chemical potential in units of $e^{2}/h$. }
\label{figure5}
\end{figure*}

Figures \ref{figure3}, \ref{figure4} and \ref{figure5} show the GGA+$U$+SOC band structures, the Berry curvatures (BCs) and anomalous Hall conductivities of $X=$ Ti, Mn and Co for in- and out-of-plane magnetization direction with P3 symmetry, for which we observe a significant effect of SOC. The corresponding results for P1 symmetry are shown in the Supplemental Material. The magnetocrystalline anisotropy and the spin and orbital moments are listed in Table~\ref{tab2}. It is noteworthy that in all cases for the high symmetry the magnetic easy axis is found to be in-plane (see Table \ref{tab2}), whereas for the systems with P1 symmetry the magnetic easy axis is out-of-plane. For $X=$ Ti SOC has only a small effect for P1 symmetry (cf. Fig. S1 in the Supplemental Material) leaving the band gap nearly unchanged for both the in-plane and out-of-plane directions (see Table \ref{tab2}). In contrast, for P3 symmetry  SOC leads to a significant band reconstruction (Fig. \ref{figure3}a, b) and a metal-to-insulator transition. The intertwined bands around the Fermi level in the absence of SOC (Fig. \ref{figure2}a) are disentangled due to SOC and split into occupied and unoccupied bands separated by a band gap of 0.83 eV for both magnetization directions. 
While the single band, comprising the valence band maximum (VBM) is mainly contributed by the central-layer, the lower two bands in the energy range between --1.3 eV and --0.3 eV exhibit a predominant interface character.  Consistent with the band occupation, a Ti$^{3+}$ valence state in all layers is obtained with SOC, reflected in similar magnetic moments 1.03/0.92/1.02 $\mu_{\rm B}$. The electron density distribution in the two energy intervals indicates a staggered $d_{xz}$ (IF), $d_{yz}$ (C) orbital polarization, similar to results for the LaTiO$_3$ bilayer \cite{DoennigDesignChernMott2016}. Concerning the topological properties of $X=$ Ti, the Berry curvature shows strong oscillations between positive and negative values, leading to a vanishing AHC for both magnetization directions (Fig. \ref{figure3}e, f).

For $X=$ Mn the band structures with P3 symmetry in the presence of SOC with [100] (in-plane) and [001] (out-of-plane) magnetization directions (Fig. \ref{figure4}a, b) exhibit no pronounced modification, compared to those in the absence of SOC (Fig. \ref{figure2}c). Still, a closer inspection reveals that SOC lifts the quadratic band touching at $\Gamma$ for both magnetization directions (Fig. \ref{figure4}a, b), the in-plane magnetization being energetically favored by 0.34 meV (see Table \ref{tab2}). In contrast to $X=$ Ti, only the middle pair of bands at the Fermi level shows a distinct interface character, whereas the lowest occupied and the highest unoccupied pairs of bands have contributions from the central with some admixture of the interface layer. The electron density distribution in the three selected energy ranges gives further insight into the contribution of the different layers and suggest degenerate $e_g$ orbital occupation with some contribution of neighboring O $2p$ states.  The Berry curvature exhibits significant negative contributions (cf. Fig. \ref{figure4}c, d) due to the avoided crossings of bands along M-K and $\textrm{M}'$-$\textrm{K}'$. This results in large negative spikes in the AHC close to $\sim -3e^2/h$ (cf. Fig. \ref{figure4}e, f), indicating nontrivial pairs of bands with Chern numbers of $\pm3$ just below/above \ef. We  have explored the effect of the $U$ parameter on the band structure and topological properties for $X=$ Mn and find that the emergence of high Chern numbers is robust beyond $U\,=\,3.0$ eV up to the studied maximum value of 6 eV. For more details, see Fig. S5 in the Supplemental Material. Additionally, we investigated the relative stability between the nontrivial P3 and trivial P1 phase under strain and find that the P3 phase is stabilized under tensile strain for a lateral lattice constant  between $a=4.04$ \AA\ and $a=4.14$ \AA\ (see Fig. S6 in the Supplemental Material), corresponding to the lattice parameters of, e.g., PrScO$_3$ or LaScO$_3$~\cite{Schlom2017}.
Moreover, the topological properties of both P1 and P3 at $a=4.04$ \AA\ (cf. Fig. S7 in the Supplemental Material) were explored. In particular, the system with P3 symmetry exhibits nontrivial topological bands accompanied by a significant, nearly integer, AHC value of $\sim 0.94e^2/h$.

\begin{figure*} [htbp]
\includegraphics[width=14cm,keepaspectratio]{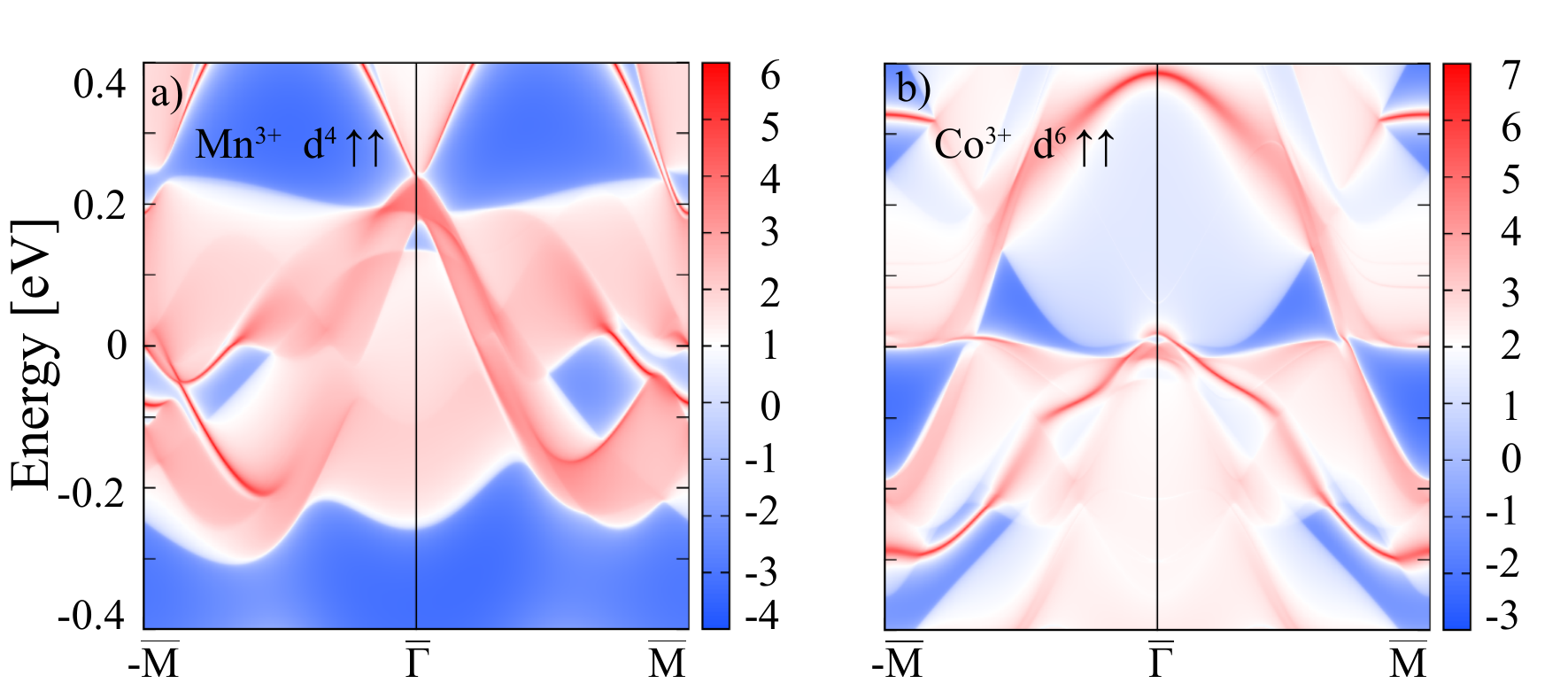}
\caption{The calculated edge states for $X$\,=\,Mn,\,Co superlattices shown in (a-b) for (100) surfaces. Red-white range of colors represent higher local DOS, the solid red lines correspond to the edge states connecting valence and conduction bands. The blue regions denote the energy gap. The Fermi level is set to zero.}
\label{figure6}
\end{figure*}

Cobalt-containing compounds tend to have a large magneto-crystalline anisotropy. This is also observed for the Co-based dice lattice which shows the highest $\Delta E$ value among the systems in the current investigation: 4.79 meV and 2.38 meV (see Table \ref{tab2}) for the P3 and P1 symmetry, respectively. Moreover, $X=$ Co acquires a substantial orbital moment of up to 0.23 and 0.24 $\mu_B$ for P3 and P1 symmetry with magnetization along [100] (cf. Table \ref{tab2}). SOC also has a stronger effect on the band structure and splits the bands at K and $\textrm{K}'$ as well as the quadratic band touching point at $\Gamma$, the effect being larger for the magnetization axis along [001] compared to the [100] crystallographic direction (see Fig. \ref{figure5}a, b). The bands around \ef\ show multiple crossings and a predominant interface character in the occupied part, mixed contribution between \ef\ and 0.8 eV, whereas the extremely flat bands above 0.8 eV show prevailing central layer contribution (see also integrated electron density distribution of the three energy regions which also indicates substantial O $2p$ hybridization). 
The larger effect of SOC for out-of-plane magnetization is also reflected in larger negative contributions to the Berry curvature $\Omega(k)$ (cf. Fig. \ref{figure5}c, d), in particular a peak arising due to the avoided crossing along $\Gamma$-M. Analogous to $X=$ Mn (cf. Fig. \ref{figure4}e) the AHC for $X=$ Co (see Fig. \ref{figure5}f) reaches values $\sim -3e^2/h$, however only for out-of-plane magnetization direction. Interestingly, the sign of the Berry curvature and Hall conductivity bear analogies to the ones identified for $X=$ Mn and Co in the (111)-oriented bilayers of La$X$O$_3$ \cite{DoennigDesignChernMott2016}. However, due to the rather semimetallic character with valence and conduction band touching close to the Fermi level the formation of a quantized Hall plateau at $E_{\rm F}$ is hampered. Nevertheless, the $\sim -3e^2/h$ peak at \ef\ indicates nontrivial bands with $C=\pm 3$. Moreover, a quantized plateau of $\sim-e^2/h$ emerges above $E_{\rm F}$ at 0.2 eV related to the avoided crossing at the K point. In order to confirm the topological character for $X=$ Mn and Co, we carried out edge state calculations by constructing the MLWFs. The edge Green's function and the local density of states (LDOS) can be simulated using an iterative method \cite{wanniertools,Sancho1984,Sancho1985}. As displayed in Fig. \ref{figure6}a, b the topologically protected chiral edge states obtained for $X=$ Mn, Co show that valence and conduction bands are connected but the edge states are obscured due to the crossing of bands at $E_{\rm F}$ in both cases.

Previous model Hamiltonian studies indicate that the dice lattice can host topological bands with nonzero AHC around the Fermi energy \cite{WangNearlyflatband2011, SoniFlatbandsferrimagnetic2020}. Our DFT+$U$+SOC calculations also predict nontrivial electronic bands with finite AHC around the Fermi energy. While Wang and Ran \cite{WangNearlyflatband2011} found bands with Chern numbers $C=\pm2$, our DFT+$U$ results indicate $C=\pm3$ for $X=$ Mn and Co. The difference can be attributed to the assumptions in the model study of an $s$-orbital and a Rashba-type SOC, while in our study the SOC effect is not related to a breaking of inversion symmetry. Moreover, the active orbital is a $3d$ orbital (\eg). The DFT+$U$ results allow to gain insight into the orbital character and address the role of lattice symmetry, atomic relaxation, orbital hybridization and spin orientation. Disentangling these aspects is pivotal to understand the electronic and topological properties of perovskite-derived dice lattices.

\section{Summary}

We have performed a systematic DFT+$U$+SOC study of the electronic and topological properties of (La$X$O$_3$)$_3$/(LaAlO$_3$)$_3$(111) SLs for $X=$ Ti, Mn and Co, where the La$X$O$_3$ (L$X$O) trilayer defines a dice lattice confined by the band insulator LaAlO$_3$. By considering lattice symmetry and SOC, we found that: (i) In the absence of SOC, when the symmetry of the three $X$ sublattices is constrained to P3, the FM phase of the L$X$O trilayer exhibits a set of spin-polarized bands with predominand interface character around the Fermi energy, leading to a distinct half-metallic state with multiple Dirac crossings and coupled electron-hole pockets; (ii) By releasing the sublattice symmetry, the FM phase of the L$X$O trilayer undergoes a significant electronic reconstruction and a metal-to-insulator transition, due to a Jahn-Teller effect ($X=$Mn) and  accompanied by a modulation of magnetic moments for $X=$ Co; (iii) SOC for P3 symmetry leads to substantial band reconstruction for $X=$ Ti resulting in a metal-to-insulator transition. While the effects of SOC are more subtle for $X=$ Mn and Co, avoided crossings close to \ef\ and lifting of the quadratic band touching at $\Gamma$ lead to anomalous Hall conductivities reaching $\sim -3e^2/h$; (iv) The band structure, anomalous Hall conductivity and Berry curvature depend strongly on the sublattice symmetry and the magnetization direction, thereby providing several degrees of freedom to tune the electronic and topological properties of perovskite-based dice lattices. While in the studied Mn and Co-based dice lattices a quantized AHC is hampered due to the (semi-) metallic character, robust Chern insulators may be achieved as a function of strain or other choice of $X$. Furthermore, electrostatic doping may be used to tune the Fermi level to the topologically nontrivial bands which can be achieved, e.g., by using polar oxide surfaces \cite{Baidya2016} or semiconductor interfaces \cite{Miao2012, Zhang2013}. The presented results for (La$X$O$_3$)$_3$/(LaAlO$_3$)$_3$(111) indicate that the dice lattice establishes a promising and rich playground to achieve exotic electronic and topological states beyond the honeycomb lattice. 

\section*{AUTHOR CONTRIBUTIONS}

The project was conceived and supervised by R. Pentcheva. Initial DFT+$U$ calculations were carried out by L. L. Li and the project was completed by O. K\"oksal. All authors contributed to the analysis and wrote the manuscript.

\section*{DATA AVAILABILITY}

The authors declare that the main data supporting the finding of this study are available within the article and its Supplementary Information files. Additional data can be provided upon request.

\section*{ACKNOWLEDGMENTS}

This work was supported by the German Science Foundation (DFG) within SFB/TRR80 (project number 107745057) project G3 and computational time at the Leibniz Supercomputer Center (project pr87ro).

\section*{COMPETING INTERESTS}

The authors declare no competing financial or non-financial interests. 

\bibliographystyle{apsrev4-2}
\bibliography{main}

\end{document}